\documentclass[manuscript]{aastex6}
\newcommand{\wind}{\textit{Wind}}
\newcommand{\insitu}{\textit{in situ}}

\usepackage{amsmath}
\usepackage{array}
\usepackage{tabulary}
\usepackage{rotating}

\usepackage{graphicx}

\shorttitle{INTERACTION BETWEEN AN IP SHOCK AND THE ASSOCIATED ICME}
\shortauthors{Liu et al.}
\begin{document}
\title{Kinetic Properties of An Interplanetary Shock Propagating Inside a Coronal Mass Ejection}
\author{Mingzhe Liu\altaffilmark{1,2}, Ying D. Liu\altaffilmark{1,2},
        Zhongwei Yang\altaffilmark{1}, L. B. Wilson III\altaffilmark{3}, and Huidong Hu\altaffilmark{1,2}}
\altaffiltext{1}{State Key Laboratory of Space Weather,
        National Space Science Center,
        Chinese Academy of Sciences, Beijing 100190, China;
        \href{mailto:liuxying@swl.ac.cn}{liuxying@swl.ac.cn};}
\altaffiltext{2}{University of Chinese Academy of Sciences, Beijing 100049, China}
\altaffiltext{3}{NASA Goddard Space Flight Center, Code 672, Greenbelt, Maryland, MD20707, USA}

\begin{abstract}
We investigate the kinetic properties of a typical fast-mode shock inside an interplanetary coronal mass ejection (ICME) observed on 1998 August 6 at 1 AU, including particle distributions and wave analysis with the \insitu{} measurements from \wind{}. Key results are obtained concerning the shock and the shock-ICME interaction at kinetic scales: (1) gyrating ions, which may provide energy dissipation at the shock in addition to wave-particle interactions, are observed around the shock ramp; (2) despite the enhanced proton temperature anisotropy of the shocked plasma, the low plasma $\beta$ inside the ICME constrains the shocked plasma under the thresholds of the ion cyclotron and mirror-mode instabilities; (3) whistler heat flux instabilities, which can pitch--angle scatter halo electrons through a cyclotron resonance, are observed around the shock, and can explain the disappearance of bidirectional electrons inside the ICME together with normal betatron acceleration; (4) whistler waves near the shock are likely associated with the whistler heat flux instabilities excited at the shock ramp, which is consistent with the result that the waves may originate from the shock ramp; (5) the whistlers share a similar characteristic with the shocklet whistlers observed by Wilson et al, providing possible evidence that the shock is decaying because of the strong magnetic field inside the ICME.

\end{abstract}


\section{Introduction}\label{1}

Interplanetary (IP) shocks are driven by coronal mass ejections (CMEs) or fast solar wind. Interactions between CMEs often result in an interesting phenomenon: a shock overtakes and penetrates a preceding CME. CME--CME interactions are a frequent phenomenon near solar maximum because multiple CMEs can occur within one day \citep{2013Liu1}. Shock--CME interactions are thus also frequent. Shock--CME interactions are of significance for both space weather predictions and basic plasma physics \citep[e.g.,][]{2012Liu, 2014Liu2, 2014Liu0, 2012Mostl, 2015Lugaz}. The studies of shock--CME interactions at kinetic scales, however, are very few.

Studies of shock kinetic properties mostly focus on the Earth's bow shock \citep[e.g.,][]{2014Wilsona, 2014Wilsonb, 2001Mobius, 2009Lee, 2012Parks, 2013Parks, 2017Parks, 2014Yang, 2016Yang}. There are some kinetic studies of IP shocks in the ambient solar wind \citep[e.g.,][]{2010Richardson1, 2007Wilson, 2009Wilson, 2012Wilson, 2016Blanco-Cano}, but none inside interplanetary CMEs (ICMEs). A well-known and long investigated feature of the Earth's bow shock is the waves and reflected particles dominant in the foreshock region. The statistics of \citet{2015Lugaz} suggest that most of the shocks inside ICMEs (low-$\beta$ plasma) are relatively weak, while \citet{1983Sckopke} find that leading or trailing wave trains are usually generated around low-Mach number low-$\beta$ shocks. Previous investigations of the waves associated with IP shocks in the ambient solar wind indicate that whistler waves play an important role in shock dynamics \citep[e.g.,][]{2009Wilson, 2012Wilson, 2017Wilson, 2012Kajdic, 2012Ramirez}. Theory and observations suggest that low Mach number shocks mainly rely upon wave dispersion for energy dissipation \citep{1985Kennel, 2017Wilson}. Gyrating ions, some of the particles reflected by the shock and gyrating around the local magnetic field, are thought to provide possible free energy for the waves near a supercritical shock \citep{2012Wilson}.

Owing to the strong magnetic field and low plasma $\beta$ of ICMEs, a shock inside an ICME is expected to have interesting properties at kinetic scales. Investigations of shock-ICME interactions so far have mainly focused on the large scale, but not on the kinetic scale processes, such as the particle kinetic behaviors and waves. Suprathermal electrons flowing in both directions along the magnetic field, called bi-directional electrons (BDEs), have been identified as a good indicator of an ICME \citep[e.g.,][]{1987Gosling}. A question thus arises regarding how a shock inside an ICME influences BDEs. Examining the particle populations and waves around a shock inside an ICME may provide a new perspective about shock--ICME interactions.

In this Letter, we analyze an IP shock inside an ICME with the \insitu{} measurements from \wind/MFI \citep{1995Lepping} and \wind/3DP \citep{1995Lin} at kinetic scales. The shock was propagating in an ICME at 1AU around 07:16:07 UT on 1998 August 6. Burst--mode particle data (full distribution every 3 seconds) from \wind/3DP around the shock are available except that the EESA-High particle detector only has the survey--mode data. With the high--cadence magnetic field data (11 samples/s) and the burst--mode particle data, we carry out a first kinetic investigation of the shock-inside-ICME phenomenon using the chosen case, including particle distributions (in Section \ref{2.1}) and wave analysis (in Section \ref{2.2}). The results provide new insights on the kinetic physical processes involved in shock-ICME interactions.

%------------------------------------------------
\section{Observations and Results}\label{2}

\begin{figure}
	\plotone{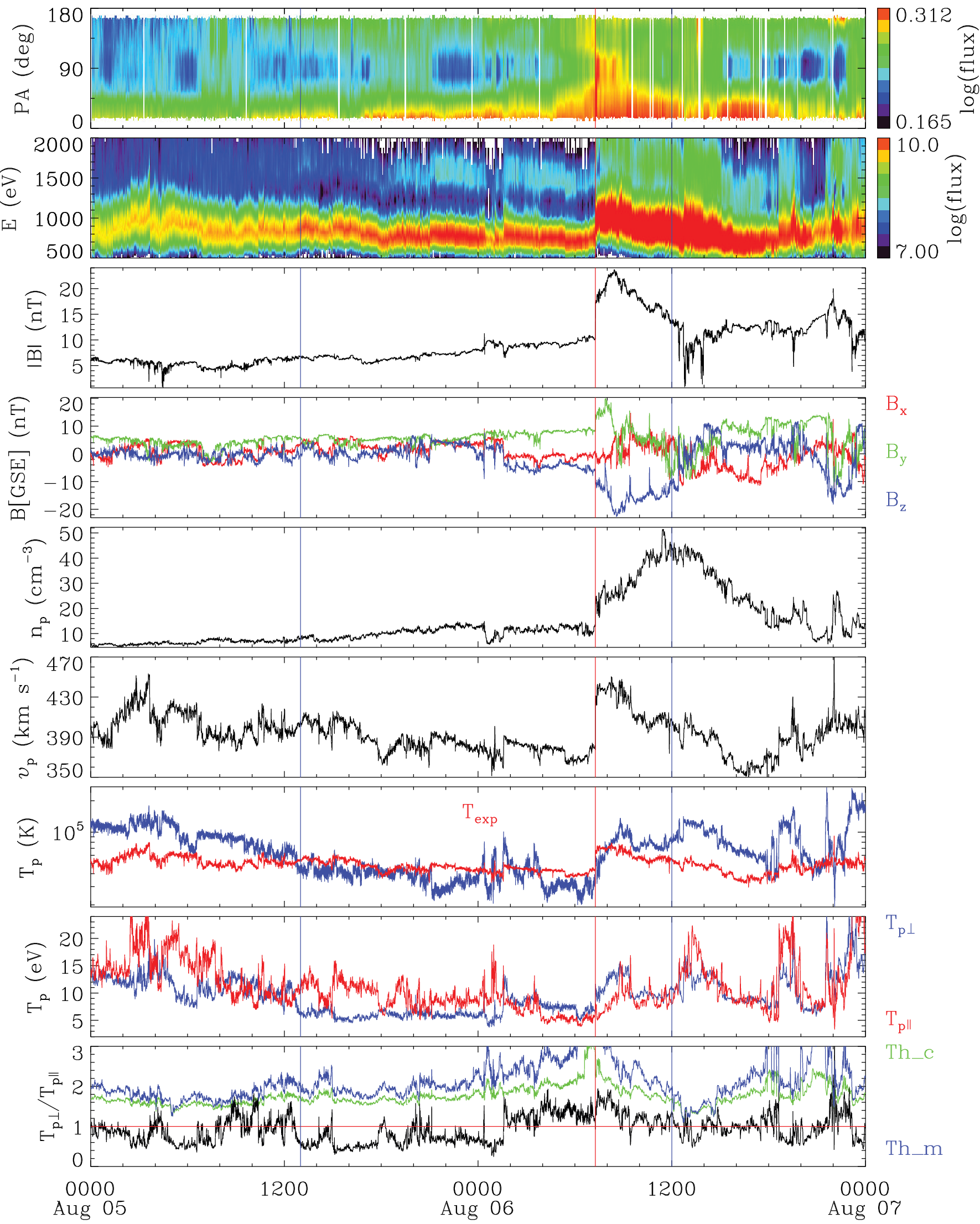}
	\caption{Solar wind measurements around the 1998 August 06 shock event from \wind{}. From top to bottom, the panels show the pitch angle distribution of 340 eV electrons, ion energy spectrum, magnetic field strength and GSE components, proton density, bulk speed, proton temperature (expected proton temperature in red), proton temperature perpendicular (blue) and parallel (red) to magnetic field, the ratio of the perpendicular temperature to the parallel temperature, and thresholds of ion cyclotron (green) and mirror-mode (blue) instabilities. The region between the two blue vertical lines is the ICME interval, and the red vertical line marks the shock propagating in the ICME.
	\label{f1}}
\end{figure}

Figure \ref{f1} shows an overview of the solar wind parameters around the IP shock observed on 1998 August 6 at \wind{}. The whole time interval of the ICME is about 23 hours. The shock has propagated deep into the ICME. The ICME shown in Figure \ref{f1} was first identified by  \citet{2010Richardson}, and then analyzed together with the shock by \citet{2015Lugaz} at large scales. The ICME is characterized by a lower-than-expected proton temperature, enhanced magnetic field strength and relatively smooth rotation of the magnetic field vector. BDEs, a typical signature of ICMEs, are observed as shown in the top panel of Figure \ref{f1}. The disappearance of BDEs after the shock may be explained partly by the pitch-angle scattering by the whistler heat flux instabilities observed near the shock ramp (see Section \ref{2.2}). The basic parameters of the shock are obtained from the shock database maintained by J. C. Kasper (https://www.cfa.harvard.edu/shocks/). The shock is a low-Mach number (${M}_{f}$ $\sim$1.59) low-$\beta$ ($\sim$ 0.07) case. It is a fast-mode quasi-perpendicular shock, with a shock normal angle ${\theta}_{Bn}$ = ${80}{\degr}$ $\pm$ ${5}{\degr}$ and a shock speed ${V}_{shn}$ = ${478.8}$ $\pm$ ${36.5}$ km s$^{-1}$.  

%------------------------------------------------
\subsection{Particle Velocity Distributions}\label{2.1}

\begin{figure*}
	\epsscale{2.0}
	\gridline{\fig{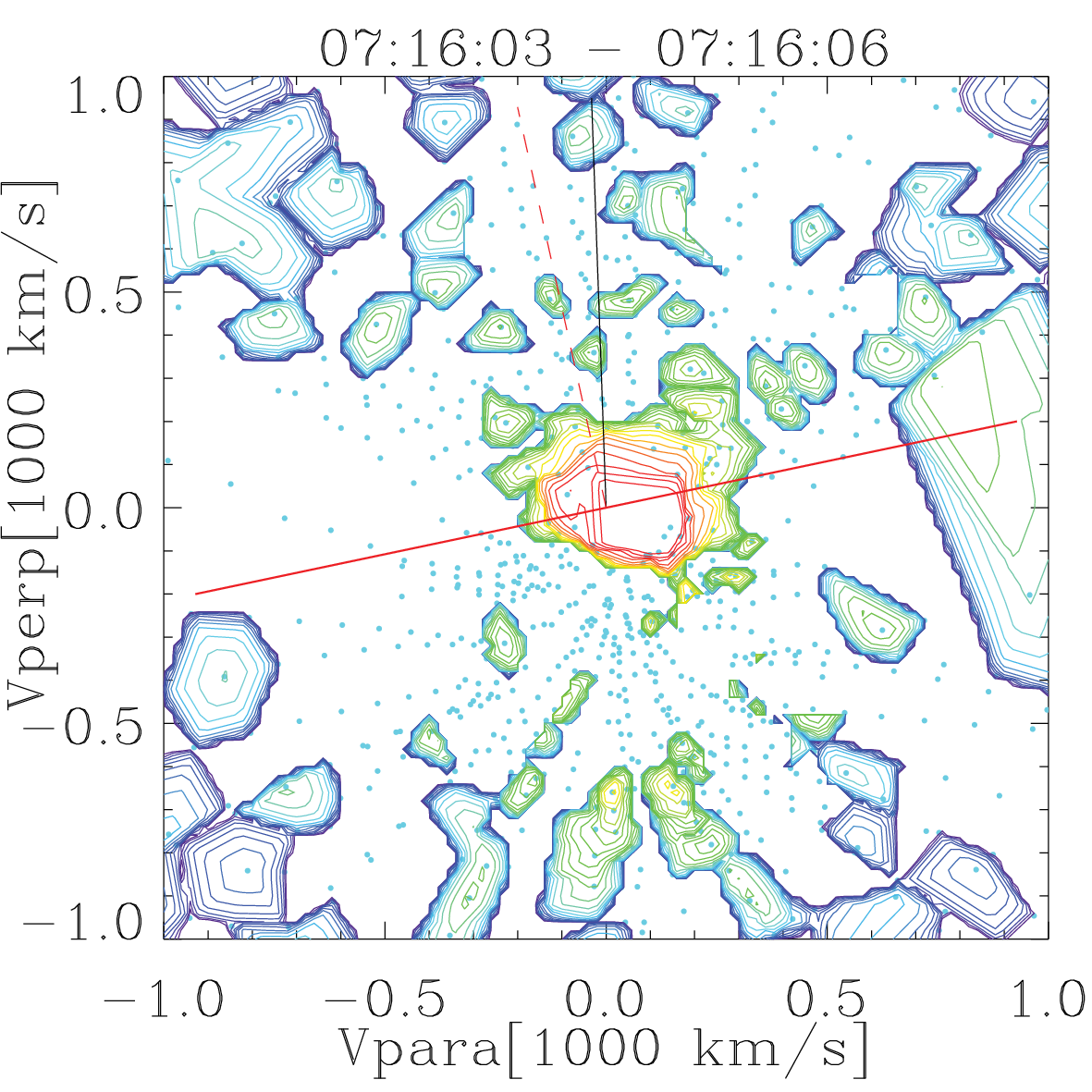}{0.333333\textwidth}{(a)upstream}
		\fig{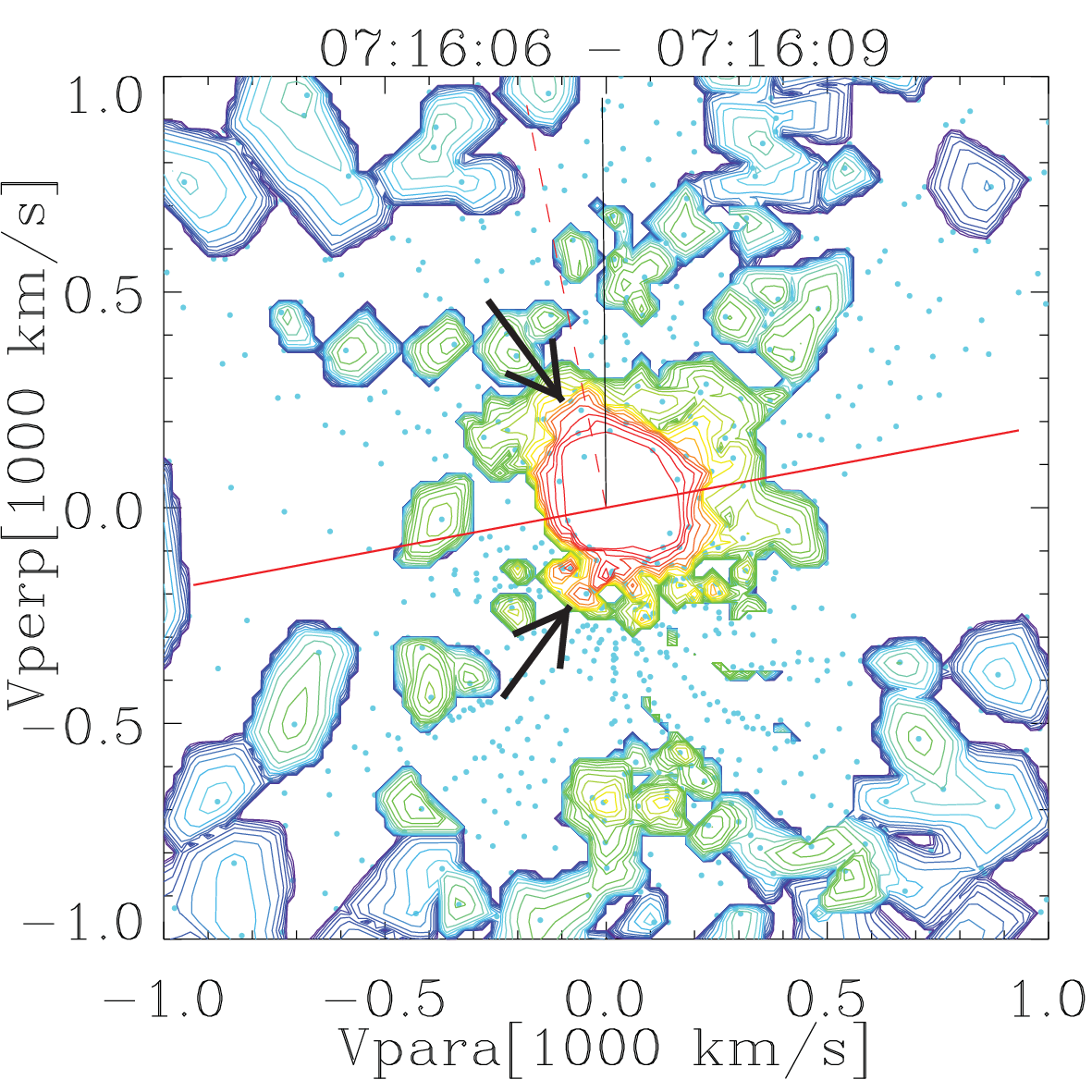}{0.333333\textwidth}{(b)foot+ramp}
		\fig{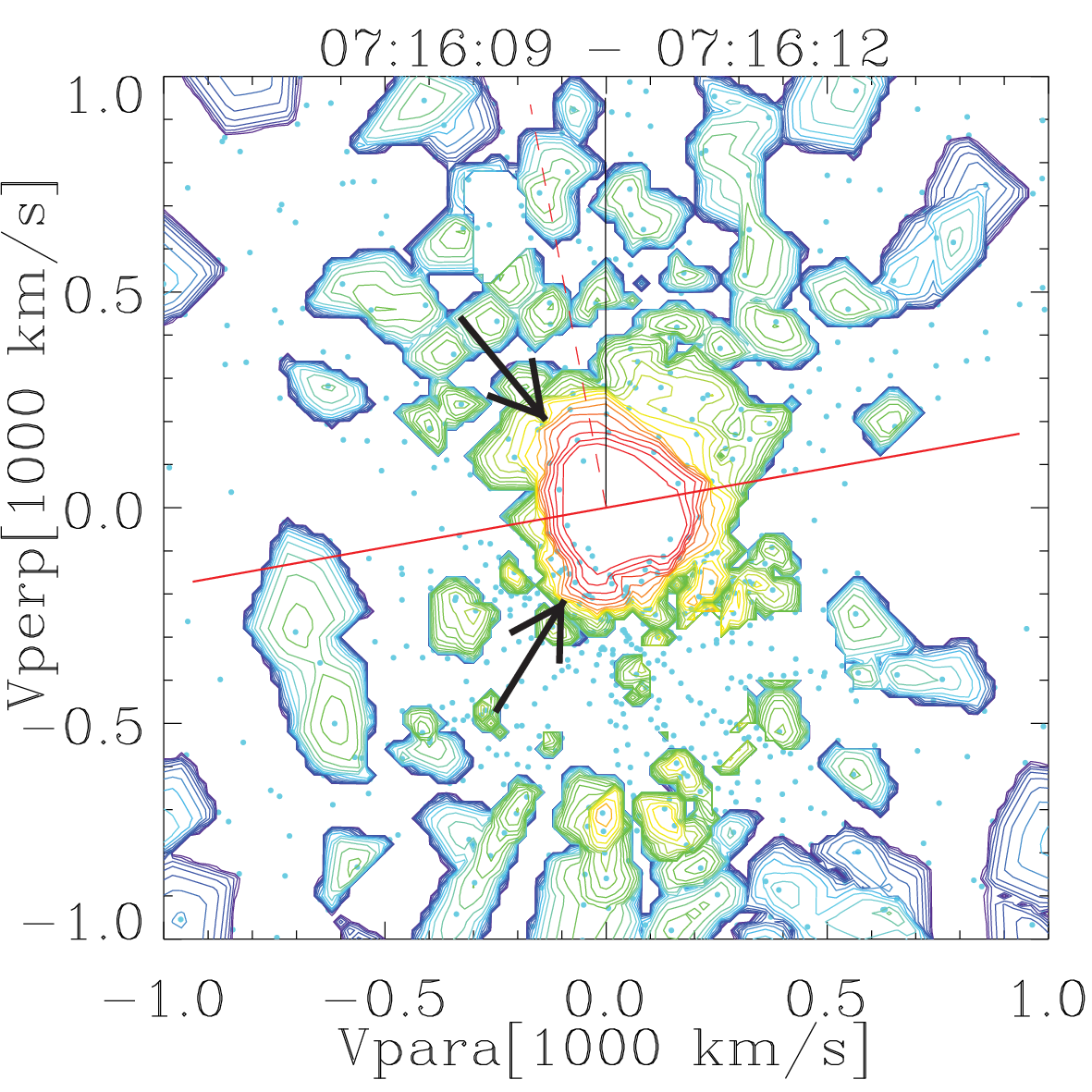}{0.333333\textwidth}{(c)downstream}
	}
	\gridline{\fig{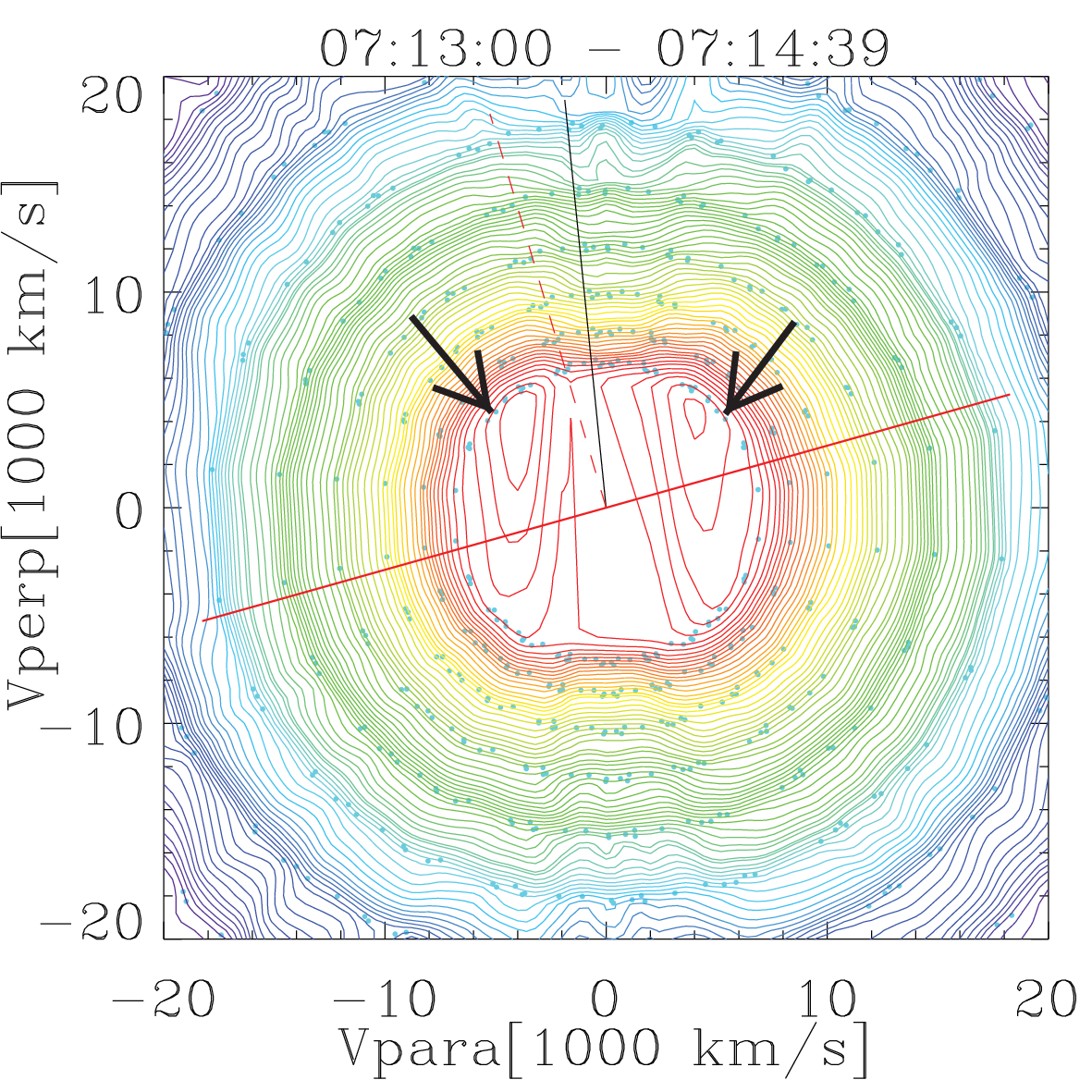}{0.333333\textwidth}{(d)upstream}
		\fig{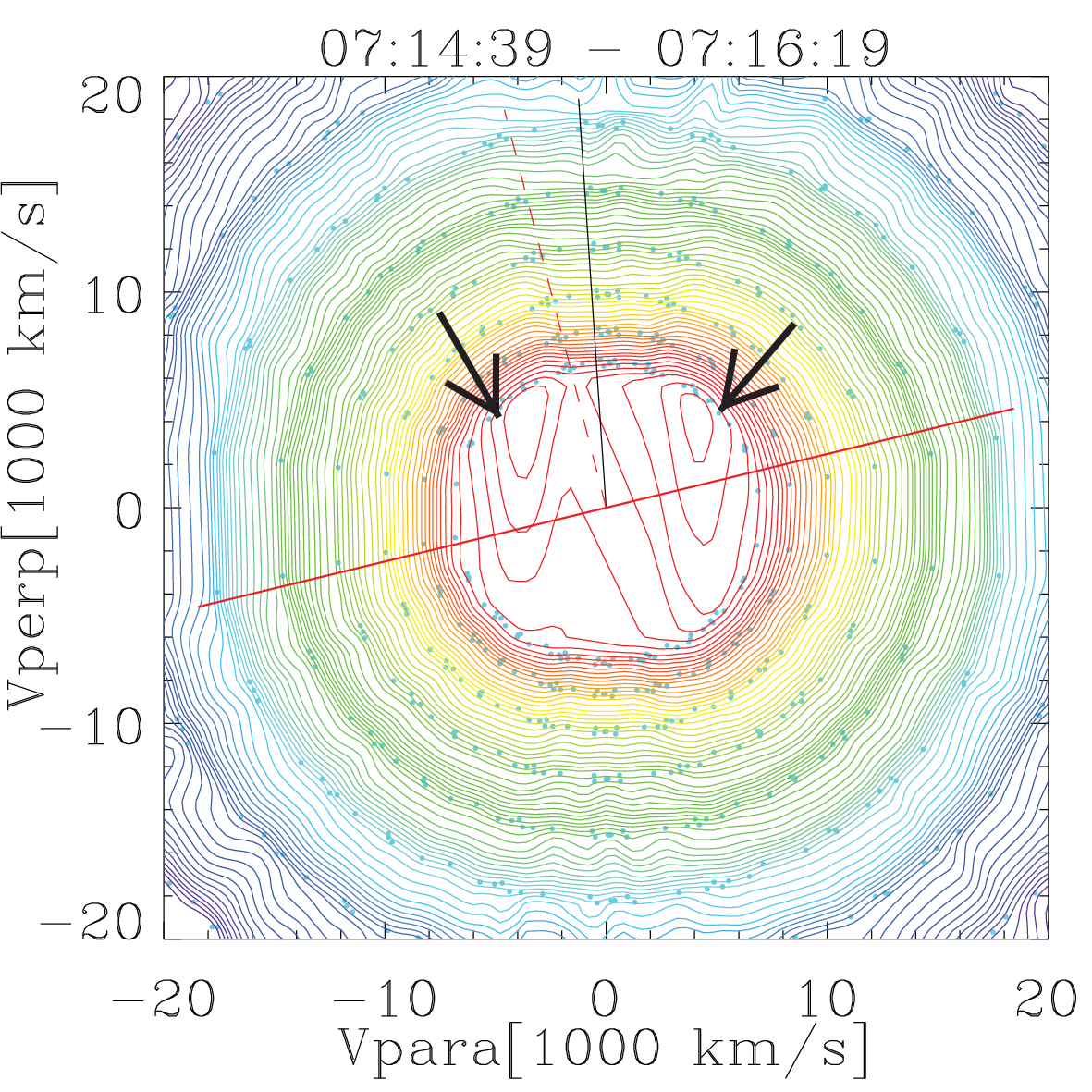}{0.333333\textwidth}{(e)upstream+ramp+downstream}
		\fig{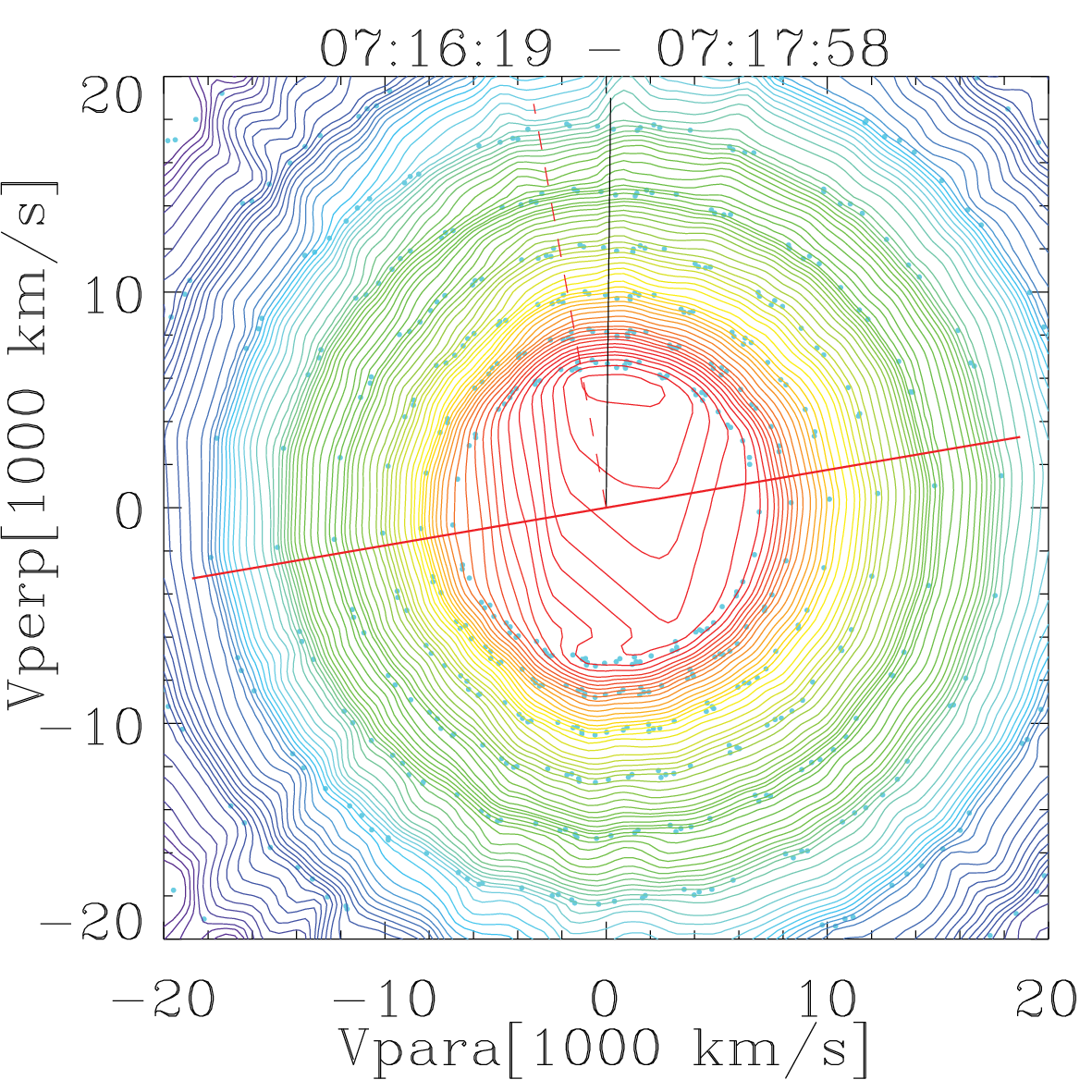}{0.333333\textwidth}{(f)downstream}
	}
	\caption{The evolution of the ion (top) and electron (bottom) distributions (in the solar wind frame) across the shock ramp. The contours show constant phase space density in the plane containing the ambient magnetic field (horizontal axis) and local solar wind velocity. Projected onto the planes are the following: shock normal direction (dashed red line), shock surface (solid red line), and solar wind velocity direction (solid black line).
	\label{f2}}
\end{figure*}

Figure \ref{f2} (top) gives the ion distributions obtained from the 3DP/PESA-High instrument around the shock ramp. The IP shock shows evidence for perpendicular ion heating, i.e., the main red beam structure is broadened much further in the direction perpendicular to the local magnetic field across the shock than in the parallel direction. A preferential perpendicular heating by the shock is also illustrated in Figure \ref{f1} by the enhanced T$_{p\bot}$/T$_{p\parallel}$ across the shock. This result is consistent with the picture of a quasi-perpendicular shock \citep[e.g.,][]{2007Liu}. \citet{2006LiuR, 2007Liu} suggest that mirror-mode instabilities resulting from high proton temperature anisotropy and high plasma $\beta$ may be excited in the downstream of a quasi-perpendicular shock. In our case, the proton temperature anisotropy of the shocked plasma is high, but the plasma $\beta$ ($\sim$0.07) is low since this is within an ICME. The low plasma $\beta$ inside the ICME inhibits mirror--mode and ion cyclotron instabilities.

We find beam-like populations of gyrating ions (indicated by black arrows) in Figure \ref{f2}b, with the velocity being about 240 km s$^{-1}$. The velocity is consistent with the estimate from a specular reflection theory \citep{1982Gosling}. The beam at the positive region of $V_{perp}$ is relatively weak.  Similarly, Figure \ref{f2}c gives evidence (indicated by black arrows) for gyrating ions in the downstream of the shock with a relatively smaller density than those in Figure \ref{f2}b. Our results on the gyrating ions are in agreement of the bow shock observations \citep{1983Sckopke} and the specular reflection theory \citep{1982Gosling}. 

Gyrating ions are usually observed in association with the magnetic foot and overshoot of quasi-perpendicular supercritical shocks due to specular reflection \citep[e.g.,][]  {1980Paschmann,1982Paschmann,1983Sckopke,1985Thomsen}.
Foot thickness roughly represents the longest distance that the specularly-reflected ions can propagate to the upstream of the shock \citep{1984Livesey, 1985Gosling}. The predicted foot thickness based on the expression from \citet{1984Livesey} is $d\sim$ 72.7 km, and the corresponding time length is $d/{V}_{p} \sim$ 0.196 s. Magnetic field observations of this shock display a transition without an obvious foot structure (see Section \ref{2.2}). The time resolution of the magnetic field data is about 11 samples/s, so very few magnetic samples are obtained within the predicted time length of the foot structure. Therefore, we consider that the lack of an obvious foot structure of the shock may be due to undersampling and may not reflect the realistic magnetic transition of the shock.
%------------------------------------------------

Figure \ref{f2} (bottom) presents the evolution of the electron distributions from the 3DP/EESA-High instrument across the shock ramp. There are obvious beam features indicated by the black arrows in Figure \ref{f2}d and Figure \ref{f2}e along the directions parallel and anti-parallel to the magnetic field. The velocity of the field-aligned beam structures ranges from 5000 km s$^{-1}$ to 6000 km s$^{-1}$, and the corresponding energy range of the field-aligned beam is approximately 278 eV-400 eV. The energy of BDEs shown in Figure \ref{f1} is 340 eV. Therefore, the field-aligned beams correspond to the BDEs observed inside the ICME. In the downstream of the shock as shown in Figure \ref{f2}f, the field-aligned beams disappear. These results are consistent with what is shown in Figure \ref{f1} (top panel).
%------------------------------------------------

\subsection{Wave Analysis}\label{2.2}

Figure \ref{3} shows the \wind{}/MFI data at the cadence of 11 samples/s across the shock, the power spectra of the magnetic field components and a minimum variance (MV) analysis example. Across the shock, the magnetic field is enhanced. This field jump lasts for $\sim$2 s, much longer than the electron cyclotron period $T_{ce}$ (0.00198 s $<$ $T_{ce}$ $<$ 0.00357 s). Therefore, the first adiabatic invariant $\mu$=E$_{\bot}$/B should be conserved during the compression, which means that normal betatron acceleration of electrons may exist. Normal betatron acceleration of electrons mainly occurs before the shock, since the magnetic field almost remains the same in the downstream of the shock. The normal betatron acceleration of electrons may contribute to the change of the pitch angle of BDEs \citep{2012Fu}. Waves are observed around the shock, and the peak frequency of the waves is 0.7 Hz  $<$ $f$ $<$ 3.0 Hz. The waves may be triggered at the shock ramp, since the amplitudes of these wave packets decrease from the shock ramp to the upstream. By comparing the first three obvious wave packets (indicated by black arrows) near the shock ramp at 1.5 $\sim$ 3.0 Hz with those at 0.7 $\sim$ 1.5 Hz, we find that the high frequency wave packets occur about 10 seconds ahead of the low frequency envelopes. This frequency dependence is consistent with that of whistler waves, whose group velocities increase with increasing frequency.

\begin{figure}[]
    \centering
	\begin{tabular}{cl}
		\includegraphics[width=0.5\textwidth]{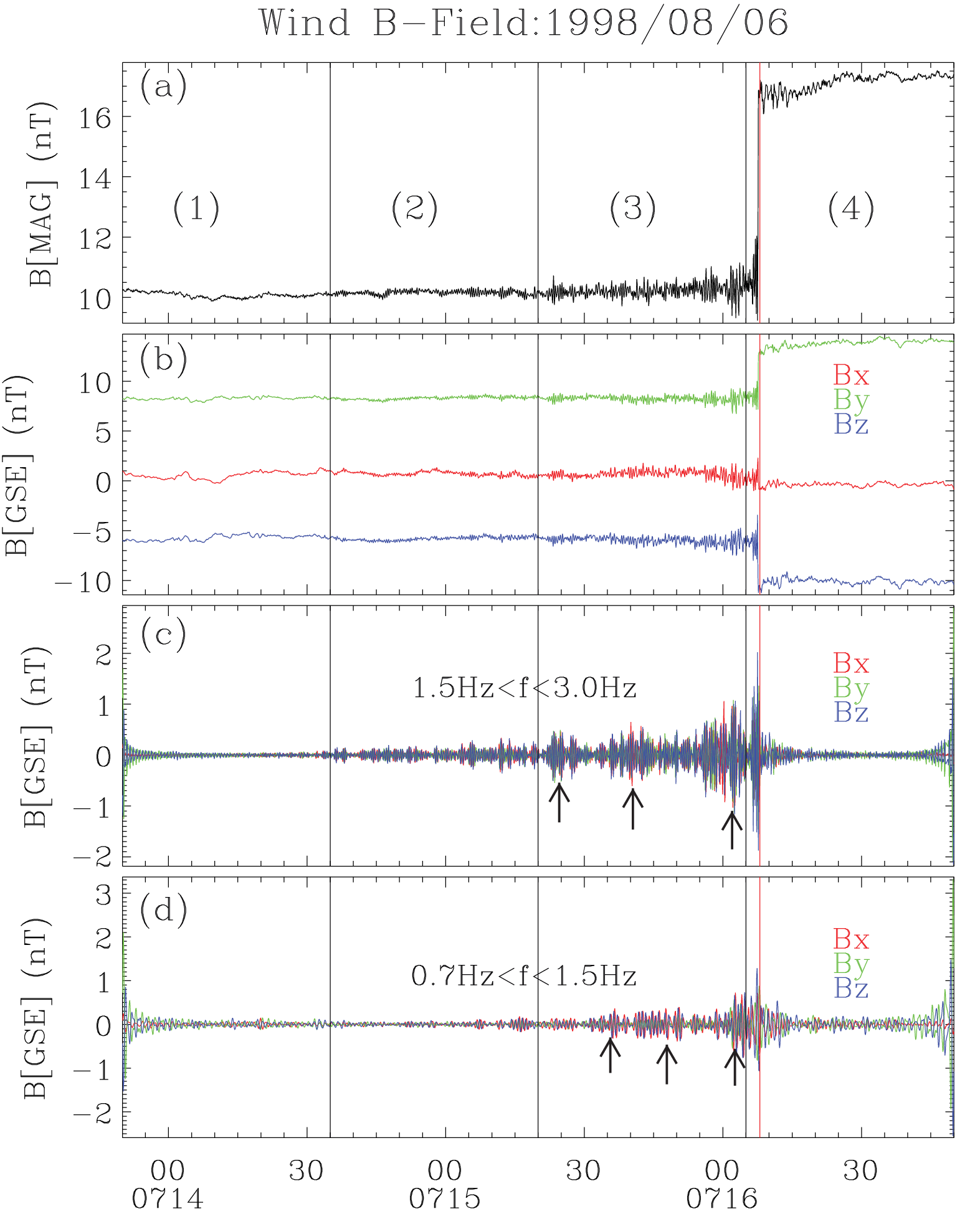} 
		\includegraphics[width=0.5\textwidth]{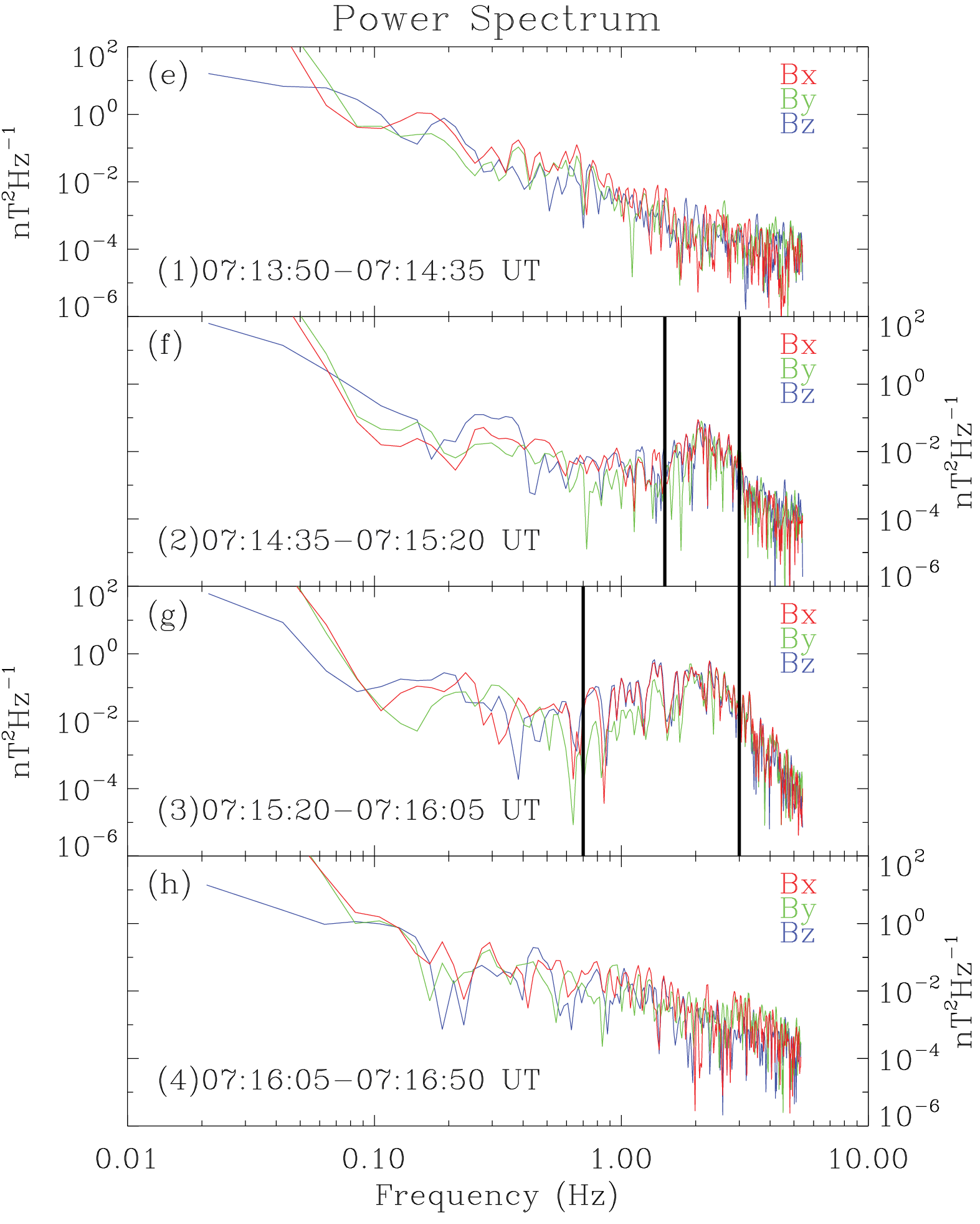} \\
		\includegraphics[width=0.7\textwidth]{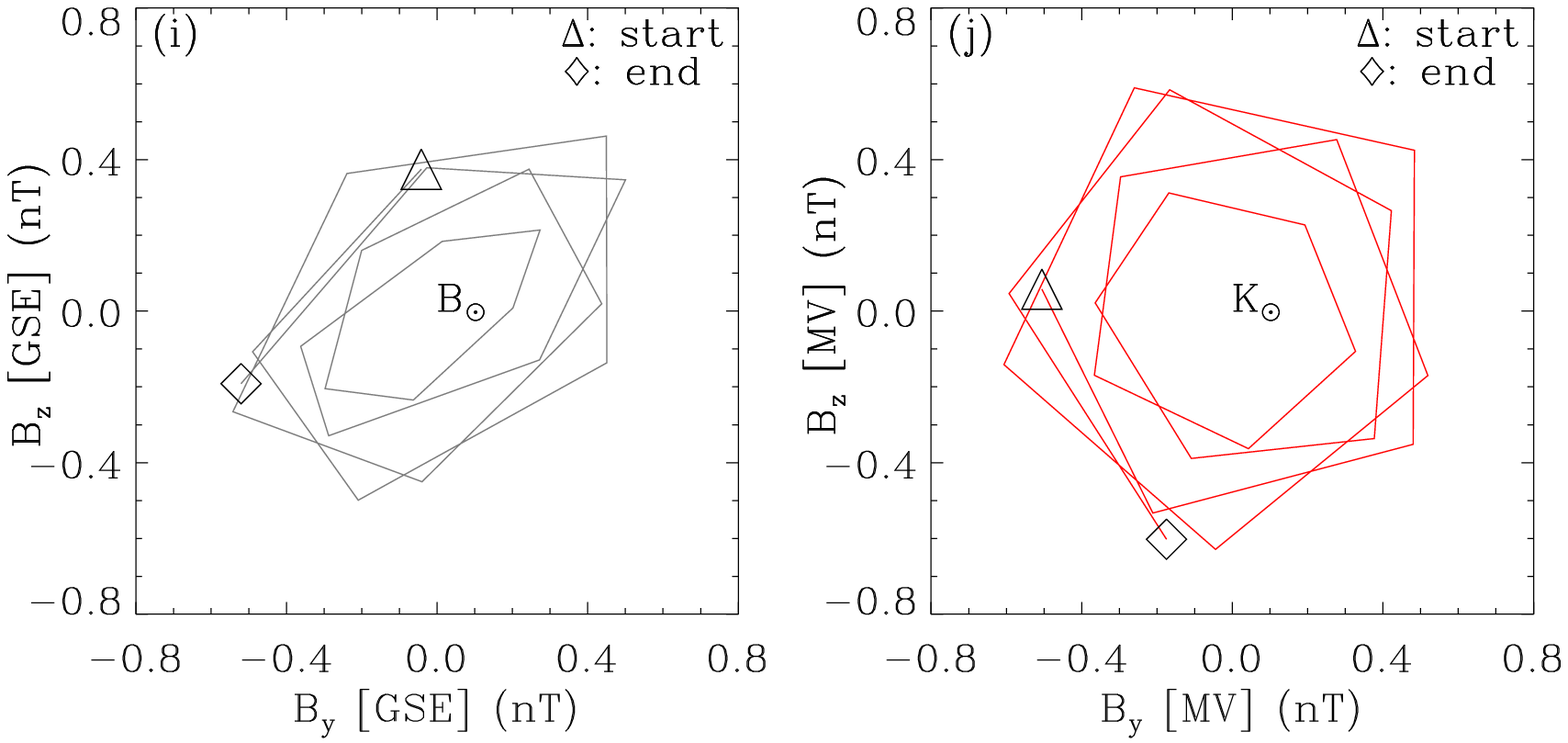} 
	\end{tabular}
	\caption{\label{f3} Magnetic field measurements and variance analysis. (a-d) Magnetic field fluctuations across the shock. From top to bottom, the panels show the magnitude of the magnetic field, the GSE components of the magnetic field, the filtered GSE components at 1.5 Hz $<$ f $<$ 3.0 Hz, and the filtered GSE components at 0.7 Hz $<$ f $<$ 1.5 Hz, respectively. The red vertical line denotes the shock arrival, and the three black vertical lines give four 45-second time intervals marked as (1), (2), (3) and (4) corresponding to the four intervals of the right panels. (e-h) Power spectra derived with a standard FFT technique. The regions between the two black vertical lines in  panels (f) and (g) indicate the peak frequency ranges of the power spectra. (i-j) An example of MV analysis of the waves (1.5Hz \textless f \textless 3.0Hz, $\lambda$$_{2}$/ $\lambda$$_{3}$=59.764, $\lambda$$_{1}$/$\lambda$$_{2}$=1.085, $\hat{K}_{GSE}$= $[$0.624, 0.487, -0.610$]$, $\theta_{kB}$=  $39^{\circ}$ (or $141^{\circ}$), $\theta_{kn}$=  $39.7^{\circ}$ (or $140.3^{\circ}$), $\theta_{kV}$=  $53.7^{\circ}$ (or $126.3^{\circ}$)) between  07:15:23 UT 
}
\end{figure}

We perform a MV analysis \citep{1998Khrabrov} of the magnetic fluctuations to determine the wave vector ($\hat{K}_{GSE}$), polarizations with respect to the ambient magnetic field and $\hat{K}_{GSE}$, and wave vector angles with respect to the local magnetic field $(\theta_{kB})$, shock normal vector $(\theta_{kn})$ and local solar wind velocity $(\theta_{kV})$. Since the upstream magnetic field and solar wind velocity are roughly constant, we use the average values obtained from the cfa shock database: $\hat{B}$=(-0.4, 8.3, -6.1) nT and $\hat{V}_{sw}$=(-371.8, -21.1, -2.9) km s${ }^{-1}$. Bandpass filters are applied to the waveforms prior to the MV analysis on specific subintervals following the procedure of \citet{2009Wilson}, which requires the ratio of the intermediate to minimum eigenvalues $\lambda$$_{2}$/$\lambda$$_{3}$ $>$10 to get a reasonable result. We choose 18 specific subintervals from 07:14:35 UT to 07:16:05 UT for the MV analysis with the frequency range: $f_{cp}$ $<$ 0.7 Hz $<$ $f$ $<$ 3.0 Hz $<$ $f_{ce}$, where $f_{cp}$ and $f_{ce}$ represent the proton gyro-frequency and electron gyro-frequency, respectively. Figure \ref{f3}i and Figure \ref{f3}j display hodograms of the magnetic field in GSE and MV coordinates. Since the X component of the magnetic field is almost zero (refer to Figure \ref{f3}b), only the hodogram with $B_z$ versus $B_y$ shows the meaningful polarization with respect to the magnetic field. The wave event is right-hand (RH) polarized (Figure \ref{f3}i) with respect to the magnetic field in the spacecraft frame. Also, the wave event is nearly circularly RH polarized with respect to $\hat{K}_{GSE}$.

 In total, the wave polarizations for the 18 subintervals with respect to $\hat{K}_{GSE}$ are: 7 samples exhibit LH sense and 11 samples present RH sense. The different polarizations of the wave events with respect to $\hat{K}_{GSE}$ can be explained by the ambiguity of the sign of $\hat{K}_{GSE}$ due to projection effects, which result from single satellite magnetic field measurements \citep{1998Khrabrov}. All the waveforms show RH polarization with respect to the local magnetic field just like the example shown in Figure \ref{f3}i, which is a characteristic of whistler waves. Since the MV analysis is performed in the spacecraft frame, we first handle the Doppler effects following the method outlined in \citet{2017Wilson} and then calculate the phase velocity of the whistler waves $(V_{ph})$, based on a cold plasma assumption. We obtain that $V_{ph}$ is about 206 km s${ }^{-1}$, which is larger than the local solar wind velocity along $\hat{K}_{GSE}$ (about 185 km s${ }^{-1}$). This suggests that the waves could be detected in their true sense of polarization with respect to the local magnetic field.

 The values of $\theta_{kB}$, $\theta_{kn}$ and $\theta_{kV}$ of the waves examined here exhibit broad ranges, consistent with the theory of \citep{1983Wu} and prior observations \citep[e.g.,][]{2009Wilson,2012Wilson,2017Wilson,2012Sundkvist}. The specific ranges of the angles are : $5.6^{\circ}$$<$ $\theta_{kB}$ $<$ $71.8^{\circ}$, $31.7^{\circ}$ $<$ $\theta_{kn}$ $<$ $74.0^{\circ}$ and $31.1^{\circ}$ $<$ $\theta_{kV}$ $<$ $81.5^{\circ}$. Compared with the observations of \citet{2009Wilson,2017Wilson}, the whistler waves in this case are similar to the shocklet whistlers. Both of them have a broad range of $\theta_{kB}$ and tend to be more oblique than the precursor whistler waves. Most of the wave samples in this Letter have $31.7^{\circ}$ $<$ $\theta_{kn}$ $<$ $74.0^{\circ}$, so they are not likely phase standing \citep{1984Mellott}.
%------------------------------------------------

\begin{figure}
	\epsscale{1.15}
	\plottwo{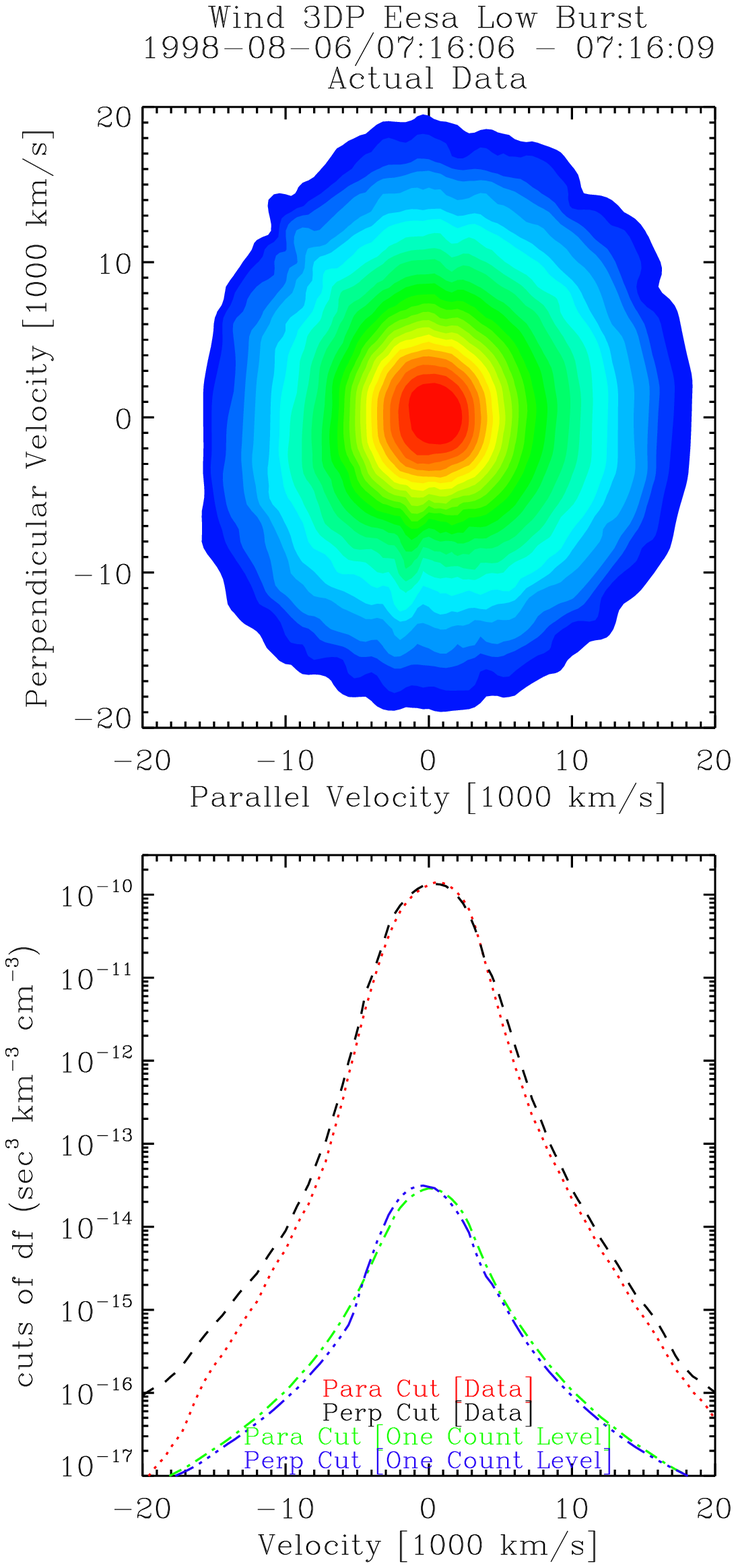}{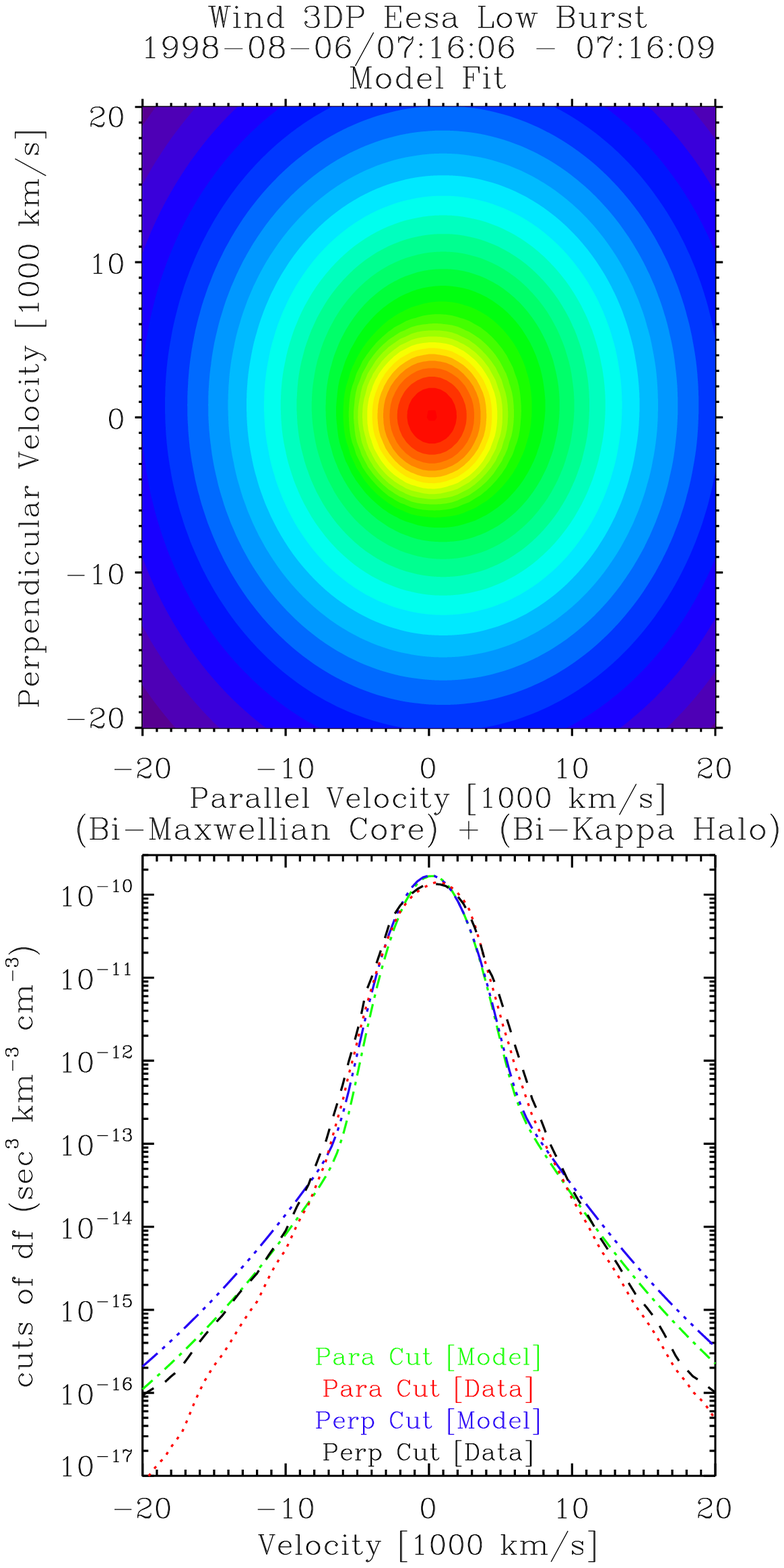}
	\caption{An example showing the comparison between the observed (left) and fitted (right) electron distributions from the 3DP/EESA-Low instrument. The top row shows the contours of the distributions. The X and Y axis denote the velocities parallel and perpendicular to the ambient magnetic field, respectively.  The bottom row presents the parallel and perpendicular cuts of the distributions and corresponding one-count level cuts. The one-count level cuts are lower than the observed value at each velocity, indicating that the measurements are reliable.
	\label{f4}}
\end{figure}

A relationship between whistler mode generation and electrons has been presented by previous studies \citep[e.g.,][]{1994Gary,1999Gary,2009Wilson}. \citet{1999Gary} demonstrate that the heat flux driven whistler mode is always unstable when the temperature anisotropy of halo electrons T$_{\bot h}$/T$_{\parallel h}$ $>$ 1.01 and always stable when the parallel beta of core electrons $\beta$$_{\parallel c}$ $\leqslant$ 0.25. The primary influence on halo electrons of whistler heat flux instabilities is to pitch angle scatter them through a cyclotron resonance. Table \ref{table1} shows statistics about the electron parameters derived from the 3DP/EESA-Low data. We fit core electrons to Bi-Maxwellian distributions and halo electrons to Bi-Kappa distributions \citep{2010Mace}. Figure \ref{f4} presents the observed and fitted electron distributions. The difference between the logarithms of the average observed and model distribution functions is less than 0.1, so the fittings are relatively good. The uncertainties of the parameters, such as the parallel/perpendicular temperatures and the number density of core/halo electrons, are in general less than 5\% of the corresponding values. From 07:16:06 UT to 07:16:09 UT, whistler heat flux instabilities arise, which is a possible driver of the whistler waves. The shock ramp is located at about 07:16:07 UT, which is near the time of the whistler heat flux instabilities, so the waves are likely produced at the shock ramp. In Table \ref{table1}, a clear increase in T$_{\bot h}$/T$_{\parallel h}$ is seen across the shock (07:16:03--07:16:09 UT), which may result from the normal cyclotron resonance that can increase the transverse energy of the electrons and normal betatron acceleration of electrons. In the downstream of the shock (07:16:07--07:16:12 UT), although normal betatron acceleration almost does not contribute to the acceleration any more, T$_{\bot h}$/T$_{\parallel h}$ still increases but at a slower rate. These results illustrate that whistler heat flux instabilities may account for the disappearance of the BDEs inside the ICME through pitch-angle scattering together with normal betatron acceleration. T$_{\bot c}$/T$_{\parallel c}$ follows the same pattern, but the increase is not so dramatic. In addition, T$_{\bot h}$/T$_{\parallel h}$ increases at a faster rate across the shock than T$_{\parallel h}$/T$_{\parallel c}$ decreases, in agreement with the observation results by \citet{2009Wilson} and simulation results by \citet{1994Gary}. Note
that there are deviations between the fit curves and the data at $V>10^4$ km s$^{-1}$. The relative deviation for the parallel cut is larger than that for the perpendicular cut, so the actual T$_{\bot h}$/T$_{\parallel h}$ should be larger than those in Table \ref{table1}. This implies that the electron distributions would be more unstable.
%------------------------------------------------

\section{Summary and Discussions}\label{3}
This Letter presents the first analysis of kinetic properties of an IP shock propagating inside an ICME on 1998 August 6, combining the particle (both ion and electron) distributions and wave analysis. Key findings are obtained concerning how shock-ICME interaction behaves at kinetic scales:
\begin{enumerate}
\item
Gyrating ions are observed around the shock, which is consistent with theoretical predictions and bow shock observations, and give evidence that particle reflection may occur even at low Mach number shocks. Reflected ions observed in this case may provide a venue for energy dissipation around the shock together with the waves observed near the shock. The shock lacks an obvious foot structure; however, the existence of gyrating ions implies that the foot structure may be under-sampled. In addition, although the shock produces enhanced proton temperature anisotropy in the downstream of the shock, the shocked ICME plasma is under the thresholds of the ion cyclotron and mirror-mode instabilities because of the low plasma $\beta$ inside the ICME.
\item
The disappearance of BDEs downstream of the shock characterizes the interaction between the shock and ICME. This is probably caused by the pitch-angle scattering of the electrons by the waves observed near the shock ramp. The electron distribution around the shock meets the criteria to excite whistler heat flux instabilities, which may contribute to the wave generation and help explain the disappearance of the BDEs downstream of the shock. Another mechanism that may also help explain the disappearance of the BDEs inside the ICME is the normal betatron acceleration that occurs across the shock.
\item
The waves around the shock are thought to be whistler waves, as the higher-frequency wave envelopes occur earlier and all the wave events show RH polarization with respect to the ambient magnetic field. The whistler waves are probably associated with the electron distribution unstable to whistler heat flux instabilities observed around the shock. The waves may originate from the shock ramp, since the amplitudes of the wave packets decrease from the shock ramp to the upstream. This is consistent with the result that the whistler heat flux instabilities are excited at the shock ramp. The whistler waves share a similar characteristic with the shocklet (steepened magnetosonic waves) whistlers, which likely suggests that the shock may be decaying due to the shock-ICME interaction. The shock has propagated deep into the ICME, so the shock may have begun to decay because of the strong magnetic field inside the ICME.
\end{enumerate}

\acknowledgments
The research was supported by the NSFC under grants 41774179, 41374173 and 41574140 and the Specialized Research Fund for State Key Laboratories of China. We acknowledge the use of data from \wind/MFI (https://spdf.sci.gsfc.nasa.gov/) and \wind/3DP    (http://sprg.ssl.berkeley.edu/).

\clearpage

\clearpage
   and  07:15:25 UT. The hodograms in GSE (i) and MV (j) coordinates are shown. The [X,Y,Z]-MV coordinates represent the directions parallel to the minimum, intermediate and maximum variance eigenvectors, respectively.

\clearpage
\begin{deluxetable}{cccccccc}
	\tablecaption{Wind 3DP electron parameters from Eesa Low Burst Mode Data \label{table1}}
	\tablehead{
		\colhead{Time(UT)} & \colhead{T$_{\bot c}$/T$_{\parallel c}$} & \colhead{T$_{\bot h}$/T$_{\parallel h}$}
		& \colhead{T$_{\parallel h}$/T$_{\parallel c}$}
		& \colhead{$\beta$$_{\parallel c}$} &
		\colhead{n$_{ce}$(cm$^{-3}$)} & \colhead{n$_{he}$(cm$^{-3}$)} &
		\colhead{n$_{he}$/n$_{ce}$}
	}
	\startdata
	07:16:00--07:16:03 & 0.93 & 0.76 & 18.01 & 0.467  & 8.66  & 0.383 & 0.044 \\
	07:16:03--07:16:06 & 0.94 & 0.77 & 18.04 & 0.469  & 8.76  & 0.387 & 0.044 \\
	07:16:06--07:16:09 & 1.08 & 1.17 & 12.69 & 0.310  & 10.47 & 0.656 & 0.063 \\
	07:16:09--07:16:12 & 1.15 & 1.47 & 10.81 & 0.202  & 11.16 & 0.786 & 0.070 \\
	07:16:12--07:16:16 & 1.16 & 1.48 & 10.80 & 0.213  & 11.12 & 0.799 & 0.072 \\
	\enddata
\end{deluxetable}
\clearpage

\clearpage
%\bibliography{references4}

\begin{thebibliography}{}
	\expandafter\ifx\csname natexlab\endcsname\relax\def\natexlab#1{#1}\fi
	
	\bibitem[{{Blanco-Cano} {et~al.}(2016){Blanco-Cano}, {Kajdi{\v c}},
		{Aguilar-Rodr{\'{\i}}guez}, {Russell}, {Jian}, \&
		{Luhmann}}]{2016Blanco-Cano}
	{Blanco-Cano}, X., {Kajdi{\v c}}, P., {Aguilar-Rodr{\'{\i}}guez}, E., {et~al.}
	2016, \jgr, 121, 992
	
	\bibitem[{{Fu} {et~al.}(2012){Fu}, {Cao}, {Mozer}, {Lu}, \& {Yang}}]{2012Fu}
	{Fu}, H.~S., {Cao}, J.~B., {Mozer}, F.~S., {Lu}, H.~Y., \& {Yang}, B. 2012,
	\jgr, 117, A01203
	
	\bibitem[{{Gary} {et~al.}(1994){Gary}, {Scime}, {Phillips}, \&
		{Feldman}}]{1994Gary}
	{Gary}, S.~P., {Scime}, E.~E., {Phillips}, J.~L., \& {Feldman}, W.~C. 1994,
	\jgr, 99, 23
	
	\bibitem[{{Gary} {et~al.}(1999){Gary}, {Skoug}, \& {Daughton}}]{1999Gary}
	{Gary}, S.~P., {Skoug}, R.~M., \& {Daughton}, W. 1999, Physics of Plasmas, 6,
	2607
	
	\bibitem[{{Gosling} {et~al.}(1987){Gosling}, {Baker}, {Bame}, {Feldman},
		{Zwickl}, \& {Smith}}]{1987Gosling}
	{Gosling}, J.~T., {Baker}, D.~N., {Bame}, S.~J., {et~al.} 1987, \jgr, 92, 8519
	
	\bibitem[{{Gosling} \& {Thomsen}(1985)}]{1985Gosling}
	{Gosling}, J.~T., \& {Thomsen}, M.~F. 1985, \jgr, 90, 9893
	
	\bibitem[{{Gosling} {et~al.}(1982){Gosling}, {Thomsen}, {Bame}, {Feldman},
		{Paschmann}, \& {Sckopke}}]{1982Gosling}
	{Gosling}, J.~T., {Thomsen}, M.~F., {Bame}, S.~J., {et~al.} 1982, \grl, 9, 1333
	
	\bibitem[{{Kajdi{\v c}} {et~al.}(2012){Kajdi{\v c}}, {Blanco-Cano},
		{Aguilar-Rodriguez}, {Russell}, {Jian}, \& {Luhmann}}]{2012Kajdic}
	{Kajdi{\v c}}, P., {Blanco-Cano}, X., {Aguilar-Rodriguez}, E., {et~al.} 2012,
	\jgr, 117, A06103
	
	\bibitem[{{Kennel} {et~al.}(1985){Kennel}, {Edmiston}, \& {Hada}}]{1985Kennel}
	{Kennel}, C.~F., {Edmiston}, J.~P., \& {Hada}, T. 1985, Washington DC American
	Geophysical Union Geophysical Monograph Series, 34, 1
	
	\bibitem[{{Khrabrov} \& {Sonnerup}(1998)}]{1998Khrabrov}
	{Khrabrov}, A.~V., \& {Sonnerup}, B.~U.~{\"O}. 1998, \jgr, 103, 6641
	
	\bibitem[{{Lee} {et~al.}(2009){Lee}, {Parks}, {Wilber}, \& {Lin}}]{2009Lee}
	{Lee}, E., {Parks}, G.~K., {Wilber}, M., \& {Lin}, N. 2009, \prl, 103, 031101
	
	\bibitem[{{Lepping} {et~al.}(1995){Lepping}, {Ac{\~u}na}, {Burlaga}, {Farrell},
		{Slavin}, {Schatten}, {Mariani}, {Ness}, {Neubauer}, {Whang}, {Byrnes},
		{Kennon}, {Panetta}, {Scheifele}, \& {Worley}}]{1995Lepping}
	{Lepping}, R.~P., {Ac{\~u}na}, M.~H., {Burlaga}, L.~F., {et~al.} 1995, \ssr,
	71, 207
	
	\bibitem[{{Lin} {et~al.}(1995){Lin}, {Anderson}, {Ashford}, {Carlson},
		{Curtis}, {Ergun}, {Larson}, {McFadden}, {McCarthy}, {Parks}, {R{\`e}me},
		{Bosqued}, {Coutelier}, {Cotin}, {D'Uston}, {Wenzel}, {Sanderson}, {Henrion},
		{Ronnet}, \& {Paschmann}}]{1995Lin}
	{Lin}, R.~P., {Anderson}, K.~A., {Ashford}, S., {et~al.} 1995, \ssr, 71, 125
	
	\bibitem[{{Liu} {et~al.}(2007){Liu}, {Richardson}, {Belcher}, \&
		{Kasper}}]{2007Liu}
	{Liu}, Y., {Richardson}, J.~D., {Belcher}, J.~W., \& {Kasper}, J.~C. 2007,
	\apjl, 659, L65
	
	\bibitem[{{Liu} {et~al.}(2006){Liu}, {Richardson}, {Belcher}, {Kasper}, \&
		{Skoug}}]{2006LiuR}
	{Liu}, Y., {Richardson}, J.~D., {Belcher}, J.~W., {Kasper}, J.~C., \& {Skoug},
	R.~M. 2006, \jgr, 111, A09108
	
	\bibitem[{{Liu} {et~al.}(2013){Liu}, {Luhmann}, {Lugaz}, {M{\"o}stl}, {Davies},
		{Bale}, \& {Lin}}]{2013Liu1}
	{Liu}, Y.~D., {Luhmann}, J.~G., {Lugaz}, N., {et~al.} 2013, \apj, 769, 45
	
	\bibitem[{{Liu} {et~al.}(2014{\natexlab{a}}){Liu}, {Yang}, {Wang}, {Luhmann},
		{Richardson}, \& {Lugaz}}]{2014Liu2}
	{Liu}, Y.~D., {Yang}, Z., {Wang}, R., {et~al.} 2014{\natexlab{a}}, \apjl, 793,
	L41
	
	\bibitem[{{Liu} {et~al.}(2012){Liu}, {Luhmann}, {M{\"o}stl},
		{Martinez-Oliveros}, {Bale}, {Lin}, {Harrison}, {Temmer}, {Webb}, \&
		{Odstrcil}}]{2012Liu}
	{Liu}, Y.~D., {Luhmann}, J.~G., {M{\"o}stl}, C., {et~al.} 2012, \apjl, 746, L15
	
	\bibitem[{{Liu} {et~al.}(2014{\natexlab{b}}){Liu}, {Luhmann}, {Kajdi{\v c}},
		{Kilpua}, {Lugaz}, {Nitta}, {M{\"o}stl}, {Lavraud}, {Bale}, {Farrugia}, \&
		{Galvin}}]{2014Liu0}
	{Liu}, Y.~D., {Luhmann}, J.~G., {Kajdi{\v c}}, P., {et~al.} 2014{\natexlab{b}},
	Nat. Commun, 5, 3481
	
	\bibitem[{{Livesey} {et~al.}(1984){Livesey}, {Russell}, \&
		{Kennel}}]{1984Livesey}
	{Livesey}, W.~A., {Russell}, C.~T., \& {Kennel}, C.~F. 1984, \jgr, 89, 6824
	
	\bibitem[{{Lugaz} {et~al.}(2015){Lugaz}, {Farrugia}, {Smith}, \&
		{Paulson}}]{2015Lugaz}
	{Lugaz}, N., {Farrugia}, C.~J., {Smith}, C.~W., \& {Paulson}, K. 2015, \jgr,
	120, 2409
	
	\bibitem[{{Mace} \& {Sydora}(2010)}]{2010Mace}
	{Mace}, R.~L., \& {Sydora}, R.~D. 2010, \jgr, 115, A07206
	
	\bibitem[{{Mellott} \& {Greenstadt}(1984)}]{1984Mellott}
	{Mellott}, M.~M., \& {Greenstadt}, E.~W. 1984, \jgr, 89, 2151
	
	\bibitem[{{M{\"o}bius} {et~al.}(2001){M{\"o}bius}, {Kucharek}, {Mouikis},
		{Georgescu}, {Kistler}, {Popecki}, {Scholer}, {Bosqued}, {R{\`e}me},
		{Carlson}, {Klecker}, {Korth}, {Parks}, {Sauvaud}, {Balsiger},
		{Bavassano-Cattaneo}, {Dandouras}, {Dilellis}, {Eliasson}, {Formisano},
		{Horbury}, {Lennartsson}, {Lundin}, {McCarthy}, {McFadden}, \&
		{Paschmann}}]{2001Mobius}
	{M{\"o}bius}, E., {Kucharek}, H., {Mouikis}, C., {et~al.} 2001, Annales
	Geophysicae, 19, 1411
	
	\bibitem[{{M{\"o}stl} {et~al.}(2012){M{\"o}stl}, {Farrugia}, {Kilpua}, {Jian},
		{Liu}, {Eastwood}, {Harrison}, {Webb}, {Temmer}, {Odstrcil}, {Davies},
		{Rollett}, {Luhmann}, {Nitta}, {Mulligan}, {Jensen}, {Forsyth}, {Lavraud},
		{de Koning}, {Veronig}, {Galvin}, {Zhang}, \& {Anderson}}]{2012Mostl}
	{M{\"o}stl}, C., {Farrugia}, C.~J., {Kilpua}, E.~K.~J., {et~al.} 2012, \apj,
	758, 10
	
	\bibitem[{{Parks} {et~al.}(2017){Parks}, {Lee}, {Fu}, {Lin}, {Liu}, \&
		{Yang}}]{2017Parks}
	{Parks}, G.~K., {Lee}, E., {Fu}, S.~Y., {et~al.} 2017, Reviews of Modern Plasma
	Physics, 1, 1
	
	\bibitem[{{Parks} {et~al.}(2012){Parks}, {Lee}, {McCarthy}, {Goldstein}, {Fu},
		{Cao}, {Canu}, {Lin}, {Wilber}, {Dandouras}, {R{\'e}me}, \&
		{Fazakerley}}]{2012Parks}
	{Parks}, G.~K., {Lee}, E., {McCarthy}, M., {et~al.} 2012, \prl, 108, 061102
	
	\bibitem[{{Parks} {et~al.}(2013){Parks}, {Lee}, {Lin}, {Fu}, {McCarthy}, {Cao},
		{Hong}, {Liu}, {Shi}, {Goldstein}, {Canu}, {Dandouras}, \&
		{R{\`e}me}}]{2013Parks}
	{Parks}, G.~K., {Lee}, E., {Lin}, N., {et~al.} 2013, \apjl, 771, L39
	
	\bibitem[{{Paschmann} {et~al.}(1980){Paschmann}, {Sckopke}, {Asbridge}, {Bame},
		\& {Gosling}}]{1980Paschmann}
	{Paschmann}, G., {Sckopke}, N., {Asbridge}, J.~R., {Bame}, S.~J., \& {Gosling},
	J.~T. 1980, \jgr, 85, 4689
	
	\bibitem[{{Paschmann} {et~al.}(1982){Paschmann}, {Sckopke}, {Bame}, \&
		{Gosling}}]{1982Paschmann}
	{Paschmann}, G., {Sckopke}, N., {Bame}, S.~J., \& {Gosling}, J.~T. 1982, \grl,
	9, 881
	
	\bibitem[{{Ram{\'{\i}}rez V{\'e}lez} {et~al.}(2012){Ram{\'{\i}}rez V{\'e}lez},
		{Blanco-Cano}, {Aguilar-Rodriguez}, {Russell}, {Kajdi{\v c}}, {Jian}, \&
		{Luhmann}}]{2012Ramirez}
	{Ram{\'{\i}}rez V{\'e}lez}, J.~C., {Blanco-Cano}, X., {Aguilar-Rodriguez}, E.,
	{et~al.} 2012, \jgr, 117, A11103
	
	\bibitem[{{Richardson} \& {Cane}(2010)}]{2010Richardson}
	{Richardson}, I.~G., \& {Cane}, H.~V. 2010, \solphys, 264, 189
	
	\bibitem[{{Richardson}(2010)}]{2010Richardson1}
	{Richardson}, J.~D. 2010, \grl, 37, L12105
	
	\bibitem[{{Sckopke} {et~al.}(1983){Sckopke}, {Paschmann}, {Bame}, {Gosling}, \&
		{Russell}}]{1983Sckopke}
	{Sckopke}, N., {Paschmann}, G., {Bame}, S.~J., {Gosling}, J.~T., \& {Russell},
	C.~T. 1983, \jgr, 88, 6121
	
	\bibitem[{{Sundkvist} {et~al.}(2012){Sundkvist}, {Krasnoselskikh}, {Bale},
		{Schwartz}, {Soucek}, \& {Mozer}}]{2012Sundkvist}
	{Sundkvist}, D., {Krasnoselskikh}, V., {Bale}, S.~D., {et~al.} 2012, \prl, 108,
	025002
	
	\bibitem[{{Thomsen} {et~al.}(1985){Thomsen}, {Gosling}, {Bame}, \&
		{Rusell}}]{1985Thomsen}
	{Thomsen}, M.~F., {Gosling}, J.~T., {Bame}, S.~J., \& {Rusell}, C.~T. 1985,
	\jgr, 90, 267
	
	\bibitem[{{Wilson} {et~al.}(2014{\natexlab{a}}){Wilson}, {Sibeck}, {Breneman},
		{Le Contel}, {Cully}, {Turner}, {Angelopoulos}, \& {Malaspina}}]{2014Wilsona}
	{Wilson}, L.~B., {Sibeck}, D.~G., {Breneman}, A.~W., {et~al.}
	2014{\natexlab{a}}, \jgr, 119, 6455
	
	\bibitem[{{Wilson} {et~al.}(2014{\natexlab{b}}){Wilson}, {Sibeck}, {Breneman},
		{Le Contel}, {Cully}, {Turner}, {Angelopoulos}, \& {Malaspina}}]{2014Wilsonb}
	---. 2014{\natexlab{b}}, \jgr, 119, 6475
	
	\bibitem[{{Wilson} {et~al.}(2007){Wilson}, {Cattell}, {Kellogg}, {Goetz},
		{Kersten}, {Hanson}, {MacGregor}, \& {Kasper}}]{2007Wilson}
	{Wilson}, III, L.~B., {Cattell}, C., {Kellogg}, P.~J., {et~al.} 2007, \prl, 99,
	041101
	
	\bibitem[{{Wilson} {et~al.}(2009){Wilson}, {Cattell}, {Kellogg}, {Goetz},
		{Kersten}, {Kasper}, {Szabo}, \& {Meziane}}]{2009Wilson}
	{Wilson}, III, L.~B., {Cattell}, C.~A., {Kellogg}, P.~J., {et~al.} 2009, \jgr,
	114, A10106
	
	\bibitem[{{Wilson} {et~al.}(2017){Wilson}, {Koval}, {Szabo}, {Stevens},
		{Kasper}, {Cattell}, \& {Krasnoselskikh}}]{2017Wilson}
	{Wilson}, III, L.~B., {Koval}, A., {Szabo}, A., {et~al.} 2017, \jgr, 122, 9115
	
	\bibitem[{{Wilson} {et~al.}(2012){Wilson}, {Koval}, {Szabo}, {Breneman},
		{Cattell}, {Goetz}, {Kellogg}, {Kersten}, {Kasper}, {Maruca}, \&
		{Pulupa}}]{2012Wilson}
	---. 2012, \grl, 39, L08109
	
	\bibitem[{Wu {et~al.}(1983)Wu, Zhou, Tsai, Guo, Winske, \&
		Papadopoulos}]{1983Wu}
	Wu, C.~S., Zhou, Y.~M., Tsai, S., {et~al.} 1983, The Physics of Fluids, 26,
	1259
	
	\bibitem[{{Yang} {et~al.}(2016){Yang}, {Huang}, {Liu}, {Parks}, {Wang}, {Lu},
		\& {Hu}}]{2016Yang}
	{Yang}, Z., {Huang}, C., {Liu}, Y.~D., {et~al.} 2016, \apjs, 225, 13
	
	\bibitem[{{Yang} {et~al.}(2014){Yang}, {Liu}, {Parks}, {Wu}, {Huang}, {Shi},
		{Wang}, \& {Hu}}]{2014Yang}
	{Yang}, Z., {Liu}, Y.~D., {Parks}, G.~K., {et~al.} 2014, \apjl, 793, L11
	
\end{thebibliography}

\bibliographystyle{aasjournal}
\end{document}